\documentclass[12pt,draftclsnofoot, onecolumn]{IEEEtran}

\usepackage{amssymb}
\usepackage[cmex10]{amsmath}
\usepackage{stfloats}
\usepackage{graphicx}
\usepackage{subfigure}
\usepackage{tabularx}
\usepackage{epsfig,epsf,color,balance,cite}
\usepackage{verbatim}
\usepackage{url}
\usepackage{bm}

\newtheorem{theorem}{\bf Theorem}

\newtheorem{lemma}{\bf Lemma}

\usepackage{algorithm}
\usepackage{algorithmic}

\usepackage{color}
\definecolor{myc1}{rgb}{0,0,0}

\begin{document}

\title{\LARGE{Energy-Efficient Wireless Communications with Distributed Reconfigurable
Intelligent Surfaces}}

\author{
\IEEEauthorblockN{Zhaohui Yang, \IEEEmembership{Member, IEEE},
                  Mingzhe Chen, \IEEEmembership{Member, IEEE},
                  Walid Saad, \IEEEmembership{Fellow, IEEE},
                  Wei Xu, \IEEEmembership{Senior Member, IEEE},
                  Mohammad Shikh-Bahaei, \IEEEmembership{Senior Member, IEEE},
                   H. Vincent Poor, \IEEEmembership{Fellow, IEEE}, and Shuguang Cui, \IEEEmembership{Fellow, IEEE}
                   \vspace{-2em}
                  }
\thanks{Z. Yang  and M. Shikh-Bahaei are with the Centre for Telecommunications Research, Department of Engineering, King's College London, WC2R 2LS, UK, Emails: yang.zhaohui@kcl.ac.uk, m.sbahaei@kcl.ac.uk.}
\thanks{M. Chen  is with the Chinese University of Hong Kong, Shenzhen, 518172, China, and also with the Department of Electrical Engineering, Princeton University, Princeton, NJ, 08544, USA, Email: mingzhec@princeton.edu.}
\thanks{W. Saad  is with the Wireless@VT, Bradley Department of Electrical and Computer Engineering, Virginia Tech, Blacksburg, VA, 24060, USA, Email: walids@vt.edu.}
\thanks{ W. Xu is with the National Mobile Communications Research
Laboratory, Southeast University, Nanjing 210096, China,  Email: wxu@seu.edu.cn.}
\thanks{H. Vincent Poor is with the Department of Electrical Engineering, Princeton University, Princeton, NJ, 08544, USA, Email: poor@princeton.edu. }
\thanks{S. Cui is with the Shenzhen Research Institute of Big Data and School of Science and Engineering, the Chinese University of Hong Kong, Shenzhen, 518172, China, Email: robert.cui@gmail.com.}
 }

\maketitle

\begin{abstract}
This paper investigates the problem of resource allocation for a wireless communication network with distributed reconfigurable intelligent surfaces (RISs). In this network,  multiple RISs are spatially distributed to serve wireless users and  the energy efficiency of the network is maximized by dynamically controlling the  on-off status of each RIS as well as optimizing the reflection coefficients matrix of the RISs. This problem is posed as a joint optimization problem of transmit beamforming and RIS control, whose goal is to maximize the energy efficiency  under minimum rate constraints of the users. To solve this problem, two iterative algorithms are proposed for the single-user case and multi-user case. For the single-user case, the phase optimization problem is solved by using a successive convex approximation method, which admits a closed-form solution at each step. Moreover, the optimal RIS on-off status is obtained by using the dual method. For the multi-user case, a low-complexity greedy searching method is proposed to solve the RIS on-off optimization problem. Simulation results show that the proposed scheme achieves up to 33\%  and 68\%  gains in terms of the energy efficiency in both single-user and multi-user cases compared to the conventional RIS scheme and  amplify-and-forward relay scheme, respectively.
\end{abstract}

\IEEEpeerreviewmaketitle

\section{Introduction}



Driven by the rapid development of advanced multimedia applications, next-generation wireless networks must support high spectral efficiency and massive connectivity~\cite{saad2019vision}.
Due to high data rate demand and massive numbers of users, energy consumption has become a challenging problem in the design of future wireless networks \cite{chen2019joint}.
In consequence, energy efficiency, defined as the ratio of spectral efficiency over power consumption, has emerged as an important performance index for deploying green and sustainable wireless networks \cite{ngo2013energy,li2011energy,buzzi2016survey,zhang2019first,7264975}. 

Recently, reconfigurable intelligent surface (RIS)-assisted wireless communication has been proposed as a potential solution for enhancing the energy efficiency of wireless networks \cite{basar2019wireless,zhang2019capacity,8580675,pan2019intelligent2,nadeem2019large,yang2020RSMARIS}. 
An RIS is a meta-surface equipped with low-cost and passive elements that can be programmed to turn the wireless channel into a partially deterministic space.
In RIS-assisted wireless communication networks, a base station (BS) sends control signals to an RIS controller so as to optimize the properties of incident waves and improve the communication quality of users.
The RIS acts as a reflector and does not perform any digitalization operation.
Hence, if properly deployed, an RIS promises much lower energy consumption than traditional amplify-and-forward (AF) relays~\cite{hum2013reconfigurable,huang2014relay,ntontin2019reconfigurable}.
However,  the effective deployment of energy-efficient RIS systems faces several challenges ranging from performance characterization to network optimization \cite{huang2019holographic}.

A number of existing works such as in \cite{huang2018achievable,jung2018performance,jung2019optimality,pan2019intelligent,zhao2019intelligent,8743496,yu2019robust} has studied the deployment of RISs in wireless networks.
In \cite{huang2018achievable},
the downlink sum-rate of an RIS assisted wireless communication system was characterized.
An asymptotic analysis of the uplink transmission rate in an
RIS-based large antenna-array system was presented in \cite{jung2018performance}.
Then, in \cite{jung2019optimality}, the authors investigated the asymptotic optimality of the achievable rate in a downlink RIS system.
Considering energy harvesting, an RIS was invoked for enhancing the sum-rate performance of a simultaneous wireless information and power transfer aided system \cite{pan2019intelligent}.
Instead of considering the availability of instantaneous channel state information (CSI), the authors in \cite{zhao2019intelligent} proposed  a two-time-scale transmission protocol to maximize the achievable sum-rate for an RIS-assisted multi-user system. 
Taking the  secrecy into consideration, the work in \cite{8743496} investigated the problem of secrecy rate maximization of an RIS assisted multi-antenna system.
Further by considering imperfect CSI, the RIS was considered to enhance the physical layer security of a wireless channel in~\cite{yu2019robust}.
Beyond the above studies,
the use of RISs for enhanced wireless energy efficiency has been studied in \cite{8741198}.
In \cite{8741198}, the authors proposed a new approach to maximize the energy efficiency of a multi-user multiple-input single-output (MISO) system by jointly controlling the transmit power of the BS and the phase shifts of the RIS.
However, only a single RIS was considered for simplicity in \cite{8741198}. Deploying a number of low-cost power-efficient RISs in future networks can cooperatively enhance the coverage of the networks.
In particular, deploying multiple RISs in wireless networks has several advantages. First, distributed RISs
can provide robust data-transmission since different RISs can be deployed geometrically apart from each other. Meanwhile, multiple RISs can provide multiple paths of received signals, which increases the received signal strength.
To our best knowledge, this is the first work that optimizes the energy efficiency for a wireless network with multiple RISs.

The main contribution of this paper is a novel energy efficient resource allocation scheme for wireless communication networks with distributed RISs.
 Our key contributions include:
\begin{itemize}
\item
 We investigate a downlink wireless communication system with distributed RISs that can be dynamically turned on or off depending on the network requirements.
 To maximize the energy efficiency of the system,
we jointly optimize the phase shifts of all RISs, the transmit beamforming of the transmitter,
and the RIS on-off status vector.
  We formulate an optimization
problem with the objective of maximizing the energy efficiency under minimum rate constraint of users, transmit power constraint, and  unit-modulus constraint of the RIS phase shifts.
\item To maximize the energy efficiency for a single user, a suboptimal solution is obtained by using a low-complexity algorithm that iteratively solves two, joint subproblems. For the joint phase and power optimization subproblem, a suboptimal phase is obtained by using the successive convex approximation (SCA) method with low complexity, and the optimal power is subsequently obtained in closed form. For the RIS on-off optimization subproblem, the dual method is used to obtain the optimal solution.
\item To maximize the energy efficiency for multiple users, an iterative algorithm is proposed through solving the phase optimization subproblem, beamforming optimization subproblem, and RIS on-off subproblem, iteratively. For the subproblem of phase optimization or beamforming optimization, we use the SCA method to obtain a suboptimal solution. For the RIS on-off optimization subproblem,
    we propose a low-complexity search method that can determine the on-off status of all RISs. 
\end{itemize}
Simulation results show that our proposed approach can enhance the energy efficiency by up to 33\%  and
68\%  gains compared to the conventional RIS scheme and  AF relay scheme, respectively.

The rest of this paper is organized as follows.
Section \uppercase\expandafter{\romannumeral2} introduces the system model and problem formulation.
Section \uppercase\expandafter{\romannumeral3} and Section \uppercase\expandafter{\romannumeral4} provide
the energy efficiency optimization with a single user and multiple users, respectively.
Simulation results are provided in Section \uppercase\expandafter{\romannumeral5} and conclusions are given in Section \uppercase\expandafter{\romannumeral6}.

Notations: In this paper, the imaginary unit of a complex number is denoted by $j=\sqrt{-1}$.
Matrices and vectors are denoted by boldface capital and lower-case letters, respectively.
Matrix $\text{diag}(x_1,\cdots,x_N)$ denotes a diagonal matrix whose diagonal components
are $x_1,\cdots,x_N$.
The real and imaginary  parts of a complex number $x$ are denoted by $\mathcal R(x)$
and $\mathcal I(\cdot)$, respectively.
$\boldsymbol x^*$, $\boldsymbol x^T$, and $\boldsymbol x^H$ respectively denote the conjugate, transpose, and
conjugate transpose of vector $\boldsymbol x$.
$[\boldsymbol x]_{n}$ and $[\boldsymbol X]_{kn}$ denote the $n$-th and $(k,n)$-th elements of the
respective vector $\boldsymbol x$ and matrix $\boldsymbol X$.
$|\boldsymbol x|$ denotes the $\ell_2$-norm of vector $\boldsymbol x$.
The distribution of a circularly symmetric complex
Gaussian variable with mean $x$ and covariance $\sigma$ is denoted by ${\mathcal {CN}}( x, \sigma)$.
\section{System Model and Problem Formulation}

\subsection{Transmission Model}

\begin{figure}[t]
\centering
\includegraphics[width=3.7in]{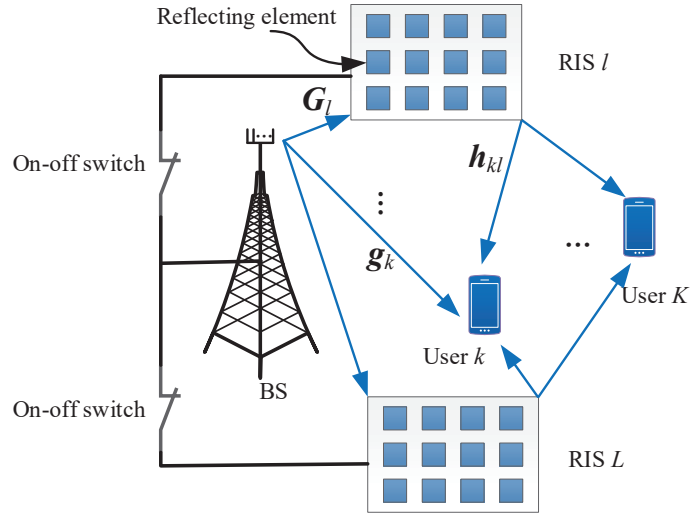}
\vspace{-2em}
\caption{A downlink  MISO system with multiple RISs.} \label{figsys1}
\vspace{-2em}
\end{figure}

Consider an RIS-assisted MISO downlink channel that consists of one BS, a set $\mathcal K$ of $K$ users, and a set $\mathcal L$ of $L$ RISs, as shown in Fig.~\ref{figsys1}.
The number of transmit antennas at the BS is $M$, while
each user is equipped with one antenna.
Such a setting has been used in many practical scenarios such as in Internet-of-Things networks \cite{huang2018achievable,jung2018performance,jung2019optimality}.
Each RIS, $l\in\mathcal L$, has $N_l$ reflecting elements.
The RISs are configured to assist the communication between the BS and users.
In particular, the RISs will be installed on the walls of the surrounding high-rise buildings.


The transmitted signal at the BS is:
\begin{equation}\label{sys2eq1}
\boldsymbol s=\sum_{k=1}^K  \boldsymbol w_k s_k,
\end{equation}
where $s_k$ is unit-power information symbol \cite{8741198} and $\boldsymbol w_k\in\mathbb C^M$ is the beamforming vector for user $k\in\mathcal K$.

The power consumption of an RIS depends on both the type and the
resolution of the reflecting elements that effectively
perform phase shifting on the impinging signal \cite{8741198,ribeiro2018energy,mendez2016hybrid}.
Considering the power consumption of RISs due to controlling the phase shift values of the reflecting elements \cite{8741198}, 
 it is often not energy efficient to turn on all the RISs.
We now introduce a binary variable $x_l\in\{0,1\}$,
where $x_l=1$ indicates that RIS $l$ is on.
When $x_l=1$, 
the phase shift matrix of RIS $l$ can be optimized through a diagonal matrix $\boldsymbol \Theta_l=\text{diag} (\text e^{j\theta_{l1}}, \cdots, \text e^{j\theta_{lN_l}})\in\mathbb C^{N_l\times  N_l}$  with $\theta_{ln}\in[0,2\pi]$, $l\in\mathcal L$, and  $n\in\mathcal N_l=\{1,\cdots,N_l\}$, where 
$\boldsymbol \Theta_l$ captures the effective phase shifts applied by all reflecting elements of RIS $l$.
In contrast, when $x_l=0$, RIS $l$ is  off and does not consume any power.
Then, with the multiple RISs, the received signal at user $k$ can be given by:
\begin{equation}\label{sys2eq2}
y_k=\left(\boldsymbol g_{k}^H+
\sum_{l=1}^L x_l\boldsymbol h_{kl}^H\boldsymbol\Theta_l  \boldsymbol G_l\right)\boldsymbol s+n_k,
\end{equation}
where $\boldsymbol g_{k}\in\mathbb C^M$,
$\boldsymbol G_l\in\mathbb C^{N_l\times M}$,  and $\boldsymbol h_{kl}\in\mathbb C^{N_l}$, respectively, denote the channel responses from the BS to user $k$, from the BS to RIS $l$, and from RIS $l$ to user $k$,
and $n_k\sim\mathcal {CN}(0,\sigma^2)$ is the additive white Gaussian noise.

Based on \eqref{sys2eq1} and \eqref{sys2eq2}, the received signal-to-interference-plus-noise ratio (SINR) at user $k$ is:
\begin{equation}\label{sys2eq3_1}
\gamma_k=\frac{ \left|\left(\boldsymbol g_{k}^H+
\sum_{l=1}^L x_l\boldsymbol h_{kl}^H \boldsymbol\Theta_l  \boldsymbol G_l\right) \boldsymbol w_k\right|^2}
{\sum_{i=1,i\neq k}^K   \left|\left(\boldsymbol g_{k}^H+
\sum_{l=1}^L x_l\boldsymbol h_{kl}^H \boldsymbol\Theta_l \boldsymbol G_l\right)\boldsymbol w_i\right|^2+\sigma^2}.
\end{equation}
As a result, the sum-rate of all users is:
\begin{equation}\label{sys2eq3_2}
R_{\text{t}}=B\sum_{k=1}^K \log_2(1+\gamma_k),
\end{equation}
where $B$ is the bandwidth of the channel.
\vspace{-.5em}
\subsection{Power Consumption Model}
\vspace{-.5em}
The total power consumption of the considered RIS-assisted system includes the transmit power of the BS, the circuit power consumption of both the BS and all users, and the power consumption of all RISs.
Consequently, the total power of the system will be given by:
\begin{equation}\label{sys2eq5}
P_{\text{t}}=  \underbrace{\sum_{k=1}^K\mu \boldsymbol w_k^H\boldsymbol w_k}_{\text{transmit power of the BS}}+ \underbrace{P_{\text{B}}}_{\text{circuit power of the BS}}+ \underbrace{\sum_{k=1}^KP_k}_{\text{circuit power of all users}}+ \underbrace{\sum_{l=1}^L x_l N_lP_{\text{R}}}_{\text{power consumption of all RISs}},
\end{equation}
where $\mu=\nu^{-1}$ with $\nu$ being the power amplifier efficiency of the BS,
$P_{\text {B}}$ is the circuit power consumption of the BS,
$P_k$ is the circuit power consumption of user $k$,
and $P_{\text{R}}$ is the power consumption of each reflecting element in the RIS.
In \eqref{sys2eq5}, $x_l N_lP_{\text{R}}$ is the power consumption of RIS $l$.

\subsection{Problem Formulation}
Given the considered system model,  our objective is to jointly optimize the reflection coefficients matrix, beamforming vector, and RIS on-off vector so as to maximize the energy efficiency 
under the minimum rate requirements and total power constraint. Mathematically, the problem for the distributed RISs  can be given by:
\begin{subequations}\label{sys2max1}
\begin{align}
\mathop{\max}_{ \boldsymbol\theta,    \boldsymbol w, \boldsymbol x} \quad& \frac{R_{\text{t}}}{P_{\text{t}}}=\frac{B\sum_{k=1}^K \log_2\left(1+\frac{ \left|\left(\boldsymbol g_{k}^H+
\sum_{l=1}^L x_l\boldsymbol h_{kl}^H \boldsymbol\Theta_l \boldsymbol G_l\right) \boldsymbol w_k\right|^2}
{\sum_{i=1,i\neq k}^K   \left|\left(\boldsymbol g_{k}^H+
\sum_{l=1}^L x_l\boldsymbol h_{kl}^H \boldsymbol\Theta_l  \boldsymbol G_l\right)\boldsymbol w_i\right|^2+\sigma^2}\right)}
{\mu \boldsymbol w^H \boldsymbol w + \sum_{k=1}^K P_k +P_{\text{B}}+\sum_{l=1}^L x_l N_lP_{\text{R}}} \tag{\theequation}  \\
\textrm{s.t.} \quad\:
&{B  \log_2 \left(1+\frac{ \left|\left(\boldsymbol g_{k}^H+
\sum_{l=1}^L x_l\boldsymbol h_{kl}^H \boldsymbol\Theta_l \boldsymbol G_l\right) \boldsymbol w_k\right|^2}
{\sum_{i=1,i\neq k}^K   \left|\left(\boldsymbol g_{k}^H+
\sum_{l=1}^L x_l\boldsymbol h_{kl}^H \boldsymbol\Theta_l  \boldsymbol G_l\right)\boldsymbol w_i\right|^2+\sigma^2}\right)}\geq R_k, \quad 
\forall k\in\mathcal K,\\
& \boldsymbol w^H \boldsymbol w \leq P_{\max},\\
&\theta_{ln} \in[0,2\pi],   \quad  \forall l \in\mathcal L, n\in\mathcal N_l, \\
& x_l\in\{0,1\}, \quad \forall  l \in\mathcal L,
\end{align}
\end{subequations}
where $\boldsymbol\theta=[\theta_{11},\cdots,\theta_{1N_1},\cdots, \theta_{LN_L}]^T$, $\boldsymbol w=[\boldsymbol w_1;\cdots; \boldsymbol w_K]$, $\boldsymbol x=[x_1,\cdots,x_L]^T$,
$R_k$ is the minimum data rate requirement of user $k$, and $P_{\max}$ is the maximum transmit power of the BS.
The minimum rate constraint  for each user is given in (\ref{sys2max1}a) and (\ref{sys2max1}b) represents the total power constraint.
The phase shift constraint for each reflecting element is provided in (\ref{sys2max1}c), which can also be seen as the unit-modulus constraint since $|\text e^{j\theta_{ln}}|=1$.
The problem in (\ref{sys2max1}) is a mixed-integer nonlinear  program  (MINLP) even for the single-user case with $K=1$.
It is generally difficult to obtain the globally optimal solution of the MINLP problem in (\ref{sys2max1}).
In the following, we propose two iterative algorithms to obtain suboptimal solutions of problem (\ref{sys2max1}) for the single-user case and the multi-user case, respectively.

\vspace{-.5em}
\section{Energy Efficiency Optimization with A Single User}\vspace{-.5em}
In this section, we consider the single-user case, i.e., $K=1$.
For $K=1$,  problem \eqref{sys2max1} becomes:
\begin{subequations}\label{sc1umax1}\vspace{-.5em}
\begin{align}
\mathop{\max}_{ \boldsymbol\theta,  \boldsymbol w_1,  \boldsymbol x} \quad& \frac{B  \log_2\left(1+\frac { \left|\left(\boldsymbol g_{1}^H+\sum_{l=1}^L x_l\boldsymbol h_{1l}^H\boldsymbol\Theta_l  \boldsymbol G_l\right) \boldsymbol w_1\right|^2}{\sigma^2}\right)}
{ \mu \boldsymbol w_1^H \boldsymbol w_1+P_1+P_{\text{B}}+\sum_{l=1}^L x_lN_lP_{\text{R}}} \tag{\theequation}  \\
\textrm{s.t.} \qquad
&{B  \log_2\left(1+\frac{ \left|\left(\boldsymbol g_{1}^H+\sum_{l=1}^L x_l\boldsymbol h_{1l}^H\boldsymbol\Theta_l  \boldsymbol G_l\right) \boldsymbol w_1\right|^2}{\sigma^2}\right)}\geq R_1, \\
 & \boldsymbol w_1^H \boldsymbol w_1 \leq P_{\max},\\
&\theta_{ln} \in[0,2\pi],   \quad  \forall l \in\mathcal L, n\in\mathcal N_l, \\
& x_l\in\{0,1\}, \quad \forall  l \in\mathcal L.
\end{align}
\end{subequations}

Since there is no multi-user interference, it is well-known that beaming as the the maximum ratio transmission (MRT) at the BS is optimal \cite{tse2005fundamentals}. That is:
\begin{equation}\label{sc1umax1eq0}
\boldsymbol w_1 = \sqrt{p_1}\frac{\boldsymbol g_{1}+\sum_{l=1}^L x_l \boldsymbol G_l^H \boldsymbol\Theta_l ^H \boldsymbol h_{1l}}
{|\boldsymbol g_{1}+\sum_{l=1}^Lx_l \boldsymbol G_l^H \boldsymbol\Theta_l ^H \boldsymbol h_{1l}|},
\end{equation}
where $p_1\leq P_{\max}$ helps satisfy the transmit power constraint at the BS.
Substituting the optimal beamforming in \eqref{sc1umax1eq0} into problem \eqref{sc1umax1}, it becomes:
\begin{subequations}\label{sc1umax2}
\begin{align}
\mathop{\max}_{ \boldsymbol\theta, p_1,  \boldsymbol x} \quad& \frac{B  \log_2\left(1+\frac{p_1\left| \boldsymbol g_{1}^H+\sum_{l=1}^L x_l \boldsymbol h_{1l}^H  \boldsymbol\Theta_l \boldsymbol G_l \right|^2}{\sigma^2}\right)}
{ \mu p_1+P_1+P_{\text{B}}+\sum_{l=1}^L x_lN_lP_{\text{R}}} \tag{\theequation}  \\
\textrm{s.t.} \quad
&{B  \log_2\left(1+\frac{p_1\left| \boldsymbol g_{1}^H+\sum_{l=1}^L x_l \boldsymbol h_{1l}^H  \boldsymbol\Theta_l \boldsymbol G_l \right|^2}{\sigma^2}\right)}\geq R_1,\\
& 0\leq p_1 \leq P_{\max}, \\
&\theta_{ln} \in[0,2\pi],   \quad  \forall l \in\mathcal L, n\in\mathcal N_l, \\
& x_l\in\{0,1\}, \quad \forall  l \in\mathcal L.
\end{align}
\end{subequations}

Due to the involvement of integer variable $\boldsymbol x$, it is difficult to obtain the globally optimal solution of \eqref{sc1umax2}.
As such, we propose an
 iterative algorithm to solve problem  \eqref{sc1umax2} sub-optimally with low complexity.
The proposed iterative algorithm contains two major steps. In the first step, we jointly optimize phase and power $( \boldsymbol\theta, p_1)$ with given $\boldsymbol x$. Then, in the second step, we update RIS on-off vector $\boldsymbol x$ with the optimized $( \boldsymbol\theta, p_1)$ in the previous step.

\subsection{Joint Phase and Power Optimization}
For a fixed integer variable $\boldsymbol x$,
problem  \eqref{sc1umax2} becomes:
\begin{subequations}\label{sc1umax2_1}
\begin{align}
\mathop{\max}_{ \boldsymbol\theta, p_1} \quad& \frac{B  \log_2\left( 1+\frac{p_1\left| \boldsymbol g_{1}^H+\sum_{l=1}^L x_l \boldsymbol h_{1l}^H  \boldsymbol\Theta_l \boldsymbol G_l \right|^2}{\sigma^2}\right)}
{ \mu p_1+P_1+P_{\text{B}}+\sum_{l=1}^L x_lN_lP_{\text{R}}} \tag{\theequation}  \\
\textrm{s.t.} \quad
&{B  \log_2\left(1+\frac{p_1\left| \boldsymbol g_{1}^H+\sum_{l=1}^L x_l \boldsymbol h_{1l}^H  \boldsymbol\Theta_l \boldsymbol G_l  \right|^2}{\sigma^2}\right)}\geq R_1,\\
& 0\leq p_1 \leq P_{\max}, \\
&\theta_{ln} \in[0,2\pi],   \quad  \forall l \in\mathcal L, n\in\mathcal N_l.
\end{align}
\end{subequations}

From the objective function (\ref{sc1umax2_1}) and the constraint in (\ref{sc1umax2_1}a), we observe that the optimal $\boldsymbol\theta$ is the one that maximizes the channel gain, i.e., $\left|\left(\boldsymbol g_{1}^H+\sum_{l=1}^L  x_l \boldsymbol h_{1l}^H  \boldsymbol\Theta_l \boldsymbol G_l\right)\right|^2$.
With this in mind, the optimal solution of problem \eqref{sc1umax2_1} can be obtained in two stages, i.e., obtain the value of $\boldsymbol\theta$ that maximizes $\left|\left(\boldsymbol g_{1}^H+\sum_{l=1}^L  x_l \boldsymbol h_{1l}^H  \boldsymbol\Theta_l \boldsymbol G_l\right)\right|^2$ in the first stage and, then, calculate the optimal $p_1$ in the second stage with the obtained $\boldsymbol \theta$ in the first stage. 

\subsubsection{First stage} We first optimize the phase shift vector $\boldsymbol\theta$ of problem \eqref{sc1umax2_1}.
Before optimizing $\boldsymbol\theta$, we show that
$\boldsymbol h_{1l}^H \boldsymbol\Theta_l \boldsymbol G_l=\boldsymbol \theta_l^T  \boldsymbol U_{1l}$, where $\boldsymbol U_{1l}=\text{diag}(\boldsymbol h_{1l}^H ) \boldsymbol G_l \in \mathbb C^{N_l\times M}$ and $\boldsymbol \theta_l=[\text e^{j\theta_{l1}}, \cdots, \text e^{j\theta_{lN_l}}]^T$.
According to problem \eqref{sc1umax2_1}, the optimal $\boldsymbol\theta$ can be calculated by solving the following problem:
\begin{subequations}\label{sc1umax3}
\begin{align}
\mathop{\max}_{ \boldsymbol\theta} \quad&   \left|\boldsymbol g_{1}^H+\sum_{l=1}^L  x_l\boldsymbol\theta_l^T \boldsymbol U_{1l}  \right|^2  \tag{\theequation}  \\
\textrm{s.t.} \quad
&\theta_{ln} \in[0,2\pi],   \quad  \forall l \in\mathcal L, n\in\mathcal N_l.
\end{align}
\end{subequations}

Let $\boldsymbol\theta_l^*$ be the conjugate vector of $\boldsymbol\theta_l$.
The total number of elements for all RISs is denoted by $Q=\sum_{l=1}^L N_l$.
Denote $\boldsymbol v=[\boldsymbol\theta_1^*;\cdots;\boldsymbol\theta_L^*]\in \mathbb C^{Q}$ and $\boldsymbol U_1=[x_1\boldsymbol U_{1l};\cdots;x_L\boldsymbol U_{1L}]\in \mathbb C^{Q\times M}$.
Problem \eqref{sc1umax3} can be rewritten as:
\begin{subequations}\label{sc1umax3_2}
\begin{align}
\mathop{\max}_{  {\boldsymbol v}} \quad&  
 \left| {\boldsymbol g}_{1} + {\boldsymbol U}_{1}^H {\boldsymbol v} \right|^2  \tag{\theequation}  \\
\textrm{s.t.} \quad
&|v_q|=1,   \quad  \forall q \in\mathcal Q,
\end{align}
\end{subequations}
where $\mathcal Q=\{1,\cdots,Q\}$.

%

To solve the optimization problem in \eqref{sc1umax3_2},
various methods were proposed by techniques such as semidefinite relaxation (SDR) technique \cite{8811733} and successive refinement (SR) algorithm \cite{Wu190603165}.
However, the SDR method imposes high complexity to obtain a rank-one solution and the SR algorithm requires a large number of iterations due to the need for updating the phase shifts in a one-by-one manner.
To reduce the computational complexity, we propose the SCA method to solve the phase shift optimization problem \eqref{sc1umax3_2}.

To handle the nonconvexity of objective function \eqref{sc1umax3_2}, we adopt the SCA method and, consequently, objective function \eqref{sc1umax3_2} can be approximated by:
\begin{align}\label{sc1umax3_2eq1}
&\quad
2\mathcal R(({\boldsymbol g}_{1} + {\boldsymbol U}_{1}^H {\boldsymbol v}^{(n-1)})^H{\boldsymbol U}_{1}^H {\boldsymbol v})-
\left| {\boldsymbol g}_{1} + {\boldsymbol U}_{1}^H {\boldsymbol v}^{(n-1)} \right|^2,
\end{align}
which is the first-order Taylor series of $ \left| {\boldsymbol g}_{1} + {\boldsymbol U}_{1}^H {\boldsymbol v} \right|^2 $
and the superscript $(n-1)$ represents the value of the variable at the $(n-1)$-th iteration.

With the above approximation \eqref{sc1umax3_2eq1}, the nonconvex problem (\ref{sc1umax3_2}) can be approximated by the following convex problem:
\begin{subequations}\label{sc1umax3_2_2}
\begin{align}
\mathop{\max}_{  {\boldsymbol v}} \quad&  2\mathcal R(({\boldsymbol g}_{1} + {\boldsymbol U}_{1}^H {\boldsymbol v}^{(n-1)})^H{\boldsymbol U}_{1}^H {\boldsymbol v})-
\left| {\boldsymbol g}_{1} + {\boldsymbol U}_{1}^H {\boldsymbol v}^{(n-1)} \right|^2   \tag{\theequation}\\
\textrm{s.t.} \quad
&|v_q|\leq 1,   \quad  \forall q \in\mathcal Q,
\end{align}
\end{subequations}
where constraint (\ref{sc1umax3_2}a) is temporarily  relaxed as (\ref{sc1umax3_2_2}a).
In the following lemma, we show that (\ref{sc1umax3_2_2}a) always holds with equality for the optimal solution of problem (\ref{sc1umax3_2_2}).
\begin{lemma}
The optimal solution of problem \eqref{sc1umax3_2_2} is:
\begin{equation}\label{sc1umax3_2_2eq1}
\boldsymbol v=\text e^{-j\angle({\boldsymbol U}_{1}({\boldsymbol g}_{1} + {\boldsymbol U}_{1}^H {\boldsymbol v}^{(n-1)}) )},
\end{equation}
where $\angle(\cdot)$ represents the angle vector of a vector, i.e,  $[\angle(\boldsymbol y)]_q=\arctan\frac{\mathcal I([\boldsymbol y]_q)}{\mathcal R([\boldsymbol y]_q)}$.
\end{lemma}

\itshape {Proof:}  \upshape
To maximize $({\boldsymbol g}_{1} + {\boldsymbol U}_{1}^H {\boldsymbol v}^{(n-1)})^H{\boldsymbol U}_{1}^H {\boldsymbol v}$ in \eqref{sc1umax3_2_2}, the optimal $\boldsymbol v$ should be chosen such that $[({\boldsymbol g}_{1} + {\boldsymbol U}_{1}^H {\boldsymbol v}^{(n-1)})^H{\boldsymbol U}_{1}^H]_q [{\boldsymbol v}]_q$ is a real number and $|[{\boldsymbol v}]_q|=1$ for any $q$, i.e., the optimal $\boldsymbol v$ should be given as \eqref{sc1umax3_2_2eq1}.
\hfill $\Box$

From \eqref{sc1umax3_2_2eq1} and Lemma 1, we can see that the optimal phase vector $\boldsymbol v$ should be adjusted such that the signal that goes through all RISs is aligned to be a signal vector with equal phase at each element.
We can also see that the optimal phase vector $\boldsymbol v$ is independent of the amplitude of the channel ${\boldsymbol U}_{1}({\boldsymbol g}_{1} + {\boldsymbol U}_{1}^H {\boldsymbol v}^{(n-1)} )$.

\begin{algorithm}[t]
\caption{SCA Method for Phase Optimization}
\begin{algorithmic}[1]\label{singlesosca}
\STATE Initialize $\boldsymbol  v^{(0)}$. Set iteration number $n=1$.
\REPEAT
\STATE Set $\boldsymbol v^{(n)}=\text e^{-j\angle({\boldsymbol U}_{1}({\boldsymbol g}_{1} + {\boldsymbol U}_{1}^H {\boldsymbol v}^{(n-1)}) )},
$ and   $n=n+1$.
\UNTIL the objective value (\ref{sc1umax3_2}) converges.
\STATE Output $\boldsymbol \theta=(\boldsymbol v^{(n)})^*$.
\end{algorithmic}
\end{algorithm}

The SCA algorithm for solving problem (\ref{sc1umax3_2}) is summarized in Algorithm \ref{singlesosca}.
The convergence of Algorithm \ref{singlesosca} is guaranteed by the following lemma that follows directly from \cite[Proposition~3]{7862919}:
\begin{lemma}\label{singlesoscalemma}
The  objective value (\ref{sc1umax3_2}) obtained in Algorithm \ref{singlesosca} is monotonically non-decreasing and the sequence $ \boldsymbol  v^{(n)}$  converges to a point fulfilling the Karush-Kuhn-Tucker (KKT) optimal conditions of the original nonconvex
problem (\ref{sc1umax3_2}).
\end{lemma}

Lemma \ref{singlesoscalemma} shows that Algorithm 1 will always converge to a locally optimal solution of problem~(\ref{sc1umax3_2}).

\subsubsection{Second stage}  We now obtain the optimal power allocation $p_1$.
With the obtained ${\boldsymbol\theta}$ in Algorithm 1 and defining $\bar g_1=\frac{\left|\boldsymbol g_{1}^H+\sum_{l=1}^L x_l\boldsymbol h_{1l}^H \boldsymbol\Theta_l \boldsymbol G_l\right|^2}{\sigma^2}$, problem \eqref{sc1umax2_1} reduces to:
\begin{subequations}\label{sc1umax3_5}
\begin{align}
\mathop{\max}_{ p_1} \quad& \frac{B  \log_2\left( 1+\bar g_1 p_1\right)}
{ \mu p_1+P_0} \tag{\theequation}  \\
\textrm{s.t.} \quad
& P_{\min}\leq p_1 \leq P_{\max} ,
\end{align}
\end{subequations}
where $P_0=P_1+P_{\text{B
}}+\sum_{l=1}^L x_lN_lP_{\text{R}}$,
and $P_{\min}=\frac{\left(2^{\frac{R_1}{B}}-1\right)}{\bar g_1}$.
In (\ref{sc1umax3_5}a), $P_{\min}$ is used to guarantee the minimum rate requirement for user 1.

For the energy efficiency optimization problem \eqref{sc1umax3_5}, the Dinkelbach method from \cite{dinkelbach1967nonlinear} can be used. The Dinkelbach method involves solving a series of convex subproblems, which increases the computational complexity.
However, the optimal solution of \eqref{sc1umax3_5} can be derived in closed form using the following theorem.
\begin{theorem}
The optimal transmit power of the energy efficiency maximization problem in \eqref{sc1umax3_5} is:
\begin{equation}\label{sc1umax3_5eq1}
p_1=\left[\frac{\bar g_1 P_0-\mu}{\mu\bar g_1 W\left( \frac{(\bar g_1 P_0-\mu)}{\mu\text e}
\right)}-\frac{1}{\bar g_1}\right]_{P_{\min}}^{P_{\max}},
\end{equation}
where $W(\cdot)$ is the Lambert-W function 
and $[a]_b^c=\min\{\max\{a,b\},c\}$.
\end{theorem}

\itshape {Proof:}  \upshape
The first-order derivative of the objective function \eqref{sc1umax3_5} with respect to power $p_1$ is:
\begin{equation}\label{sc1umax3_5eq1_1}
\frac{\partial \frac{B  \log_2( 1+\bar g_1 p_1)}
{ \mu p_1+P_0}}
{\partial p_1}=
 \frac{B  ( {\bar g_1} (\mu p_1+P_0)- \mu (1+\bar g_1 p_1) \ln( 1+\bar g_1 p_1 ))}
{ ( 1+\bar g_1 p_1)(\mu p_1+P_0)^2\ln2}.
\end{equation}

To show the increasing trend of the objective function \eqref{sc1umax3_5}, we further denote:
\begin{equation}\label{sc1umax3_5eq1_0}
f_1(p_1)= \bar g_1  (\mu p_1+P_0)- \mu (1+\bar g_1 p_1) \ln( 1+\bar g_1 p_1 ), \quad \forall p_1>0.
\end{equation}
The first-order derivative of function $f_1(p_1)$ is:
\begin{equation}\label{sc1umax3_5eq1_2}
 f_1'(p_1)=   - \mu  \bar g_1  \ln( 1+\bar g_1 p_1 )<0,
\end{equation}
which indicates that $f_1(p_1)$ is a monotonically decreasing function.
Since $ f_1(0)=\bar g_1  P_0>0$ and $\lim_{p_1\rightarrow \infty}f_1(p_1)<0$, there must exist a unique $\bar p_1$ such that $f_1(\bar p_1)=0$, where
\begin{equation}
\bar p_1=\frac{\bar g_1 P_0-\mu}{\mu\bar g_1 W\left( \frac{(\bar g_1 P_0-\mu)}{\mu\text e}
\right)}-\frac{1}{\bar g_1}.
\end{equation}
Hence, the objective function \eqref{sc1umax3_5} first increases in interval $(0,\bar p_1]$ and then decreases in interval $(\bar p_1,\infty)$, which indicates that
the optimal solution can be presented as in \eqref{sc1umax3_5eq1}.
 \hfill $\Box$

From Theorem 1, the optimal power control of problem  \eqref{sc1umax3_5} is obtained in closed-form, as shown in \eqref{sc1umax3_5eq1}.
According to \eqref{sc1umax3_5eq1}, it is shown that the optimal power control lies in one of three values, i.e., the minimum transmit power, the power with zero first-order derivative, and the maximum transmit power.

\subsection{RIS On-Off Optimization}

Substituting the phase and power variables $(\boldsymbol \theta, p_1)$ obtained in the previous section,
problem  \eqref{sc1umax2} becomes:
\begin{subequations}\label{sc1umax5}
\begin{align}
\mathop{\max}_{  \boldsymbol x} \quad& \frac{B  \log_2\left( 1+\frac{p_1\left| \boldsymbol g_{1}^H+\sum_{l=1}^L x_l \boldsymbol h_{1l}^H  \boldsymbol\Theta_l \boldsymbol G_l \right|^2}{\sigma^2}\right)}
{ \mu p_1+P_1+P_{\text{B}}+\sum_{l=1}^L x_lN_lP_{\text{R}}} \tag{\theequation}  \\
\textrm{s.t.} \quad
&{B  \log_2\left(1+\frac{p_1\left| \boldsymbol g_{1}^H+\sum_{l=1}^L x_l \boldsymbol h_{1l}^H  \boldsymbol\Theta_l \boldsymbol G_l \right|^2}{\sigma^2}\right)}\geq R_1,\\
& x_l\in\{0,1\}, \quad \forall  l \in\mathcal L.
\end{align}
\end{subequations}
We introduce an auxiliary variable $y$ and problem \eqref{sc1umax5} is equivalent to:
\begin{subequations}\label{sc1umax5_1}
\begin{align}
\mathop{\max}_{  \boldsymbol x,y} \quad& \frac{B  \log_2\left( 1+\frac{p_1y}{\sigma^2}\right)}{ \mu p_1+P_1+P_{\text{B}}+\sum_{l=1}^L x_lN_lP_{\text{R}}}
 \tag{\theequation}  \\
\textrm{s.t.} \quad
& y\leq { \left| \boldsymbol g_{1}^H+\sum_{l=1}^L x_l \boldsymbol h_{1l}^H  \boldsymbol\Theta_l \boldsymbol G_l \right|^2},\\
&y\geq \left(2^{\frac{R_1}{B}}-1\right)\frac{\sigma^2}{p_1}, \\
& x_l\in\{0,1\}, \quad \forall  l \in\mathcal L,
\end{align}
\end{subequations}
where constraint (\ref{sc1umax5_1}b) is used to ensure the minimum rate demand.
For the optimal solution of problem \eqref{sc1umax5_1}, constraint (\ref{sc1umax5_1}a) will always hold with equality since the objective function monotonically increases with $y$.
Although problem (\ref{sc1umax5_1}) has a simplifier form compared to (\ref{sc1umax5}), it is still a nonconvex MINLP.
There are two difficulties in solving problem (\ref{sc1umax5_1}).
 The first difficulty is that objective function \eqref{sc1umax5_1} has a fractional form, which is difficult to solve.
The second difficulty is that constraint (\ref{sc1umax5_1}a) is nonconvex.

To handle the first difficulty, we use the parametric approach in \cite{dinkelbach1967nonlinear} and consider the following problem:
\begin{equation}\label{equvialentProblem}
H(\lambda)=\mathop{\max}_{(\boldsymbol x,y)\in\mathcal F} {B  \log_2\left( 1+\frac{p_1y}{\sigma^2}\right)}-\lambda\left({ \mu p_1+P_1+P_{\text{B}}+\sum_{l=1}^L x_lN_lP_{\text{R}}}\right),
\end{equation}
where $\mathcal F$ is the feasible set of $(\boldsymbol x,y)$ satisfying constraints (\ref{sc1umax5_1}a)-(\ref{sc1umax5_1}c).
It was proved in \cite{dinkelbach1967nonlinear} that solving (\ref{sc1umax5_1}) is equivalent to finding the root of the nonlinear function $H(\lambda)$, which can be obtained by using the Dinkelbach method.
By introducing parameter $\lambda$, the objective function of problem (\ref{sc1umax5_1}) can be simplified, as shown in \eqref{equvialentProblem}.

To handle the second difficulty, due to the fact that $x_l\in\{0,1\}$,
 we can rewrite the right hand side of constraint (\ref{sc1umax5_1}a) as:
\begin{align}\label{sc1umax5_1eq1}
 \left| \boldsymbol g_{1}^H+\sum_{l=1}^L x_l \boldsymbol h_{1l}^H  \boldsymbol\Theta_l \boldsymbol G_l \right|^2
= D_{0}+\sum_{l=1}^LD_{l}x_l+ \sum_{l=2}^L\sum_{m=1}^{l-1}D_{lm} x_l x_m,
\end{align}
where $D_0=\boldsymbol g_{1}^H\boldsymbol g_{1}$, $D_l= \boldsymbol h_{1l}^H  \boldsymbol\Theta_l \boldsymbol G_l \boldsymbol G_l^H \boldsymbol\Theta_l^H\boldsymbol h_{1l}+\boldsymbol g_{1}^H\boldsymbol G_l^H \boldsymbol\Theta_l^H\boldsymbol h_{1l}+\boldsymbol h_{1l}^H  \boldsymbol\Theta_l \boldsymbol G_l\boldsymbol g_{1}$,
and $D_{lm}=\boldsymbol h_{1l}^H  \boldsymbol\Theta_l \boldsymbol G_l\boldsymbol G_m^H \boldsymbol\Theta_m^H\boldsymbol h_{1m}+\boldsymbol h_{1m}^H  \boldsymbol\Theta_m \boldsymbol G_m\boldsymbol G_l^H \boldsymbol\Theta_l^H\boldsymbol h_{1l}$.
To solve
problem (\ref{sc1umax5_1}), we introduce new variable $z_{lm}=x_lx_m$.
Since $x_{l} \in\{0,1\}$, constraint $z_{lm}=x_lx_m$ is equivalent to:
\begin{equation}\label{sc1umax5_1eq2}
z_{lm}\geq x_{l}+ x_{m}-1, \quad 0\leq z_{lm} \leq 1,
\end{equation}
\begin{equation}\label{sc1umax5_1eq3}
z_{lm}\leq x_{l} , \quad z_{lm}\leq x_{m},
\end{equation}
for all $l=2,\cdots, L$, $m=1,\cdots,l-1$.
According to \eqref{equvialentProblem}-\eqref{sc1umax5_1eq3}, problem (\ref{sc1umax5_1}) can be reformulated as:
\begin{subequations}\label{sc1umax5_0}
\begin{align}
\mathop{\max}_{  \boldsymbol x,y,\boldsymbol z} \quad&  {B  \log_2\left( 1+\frac{p_1y}{\sigma^2}\right)}-\lambda\left({ \mu p_1+P_1+P_{\text{B}}+\sum_{l=1}^L x_lN_lP_{\text{R}}}\right)
 \tag{\theequation}  \\
\textrm{s.t.} \quad
& y\leq D_{0}+\sum_{l=1}^LD_{l}x_l+ \sum_{l=2}^L\sum_{m=1}^{l-1}D_{lm} z_{lm},\\
& z_{lm}\geq x_{l}+ x_{m}-1, z_{lm}\leq x_{l} , z_{lm}\leq x_{m}, \quad \forall l=2,\cdots, L, m=1,\cdots,l-1,\\
& y\geq \left(2^{\frac{R_1}{B}}-1\right)\frac{\sigma^2}{p_1},\\
&0\leq z_{lm}\leq1, \quad \forall l=2,\cdots, L, m=1,\cdots,l-1,\\
& x_l\in\{0,1\}, \quad \forall  l \in\mathcal L,
\end{align}
\end{subequations}
where $\boldsymbol z=[z_{21},z_{31},z_{32},\cdots,z_{L(L-1)}]^T$.

Due to constraints (\ref{sc1umax5_0}e), it is difficult to handle  problem \eqref{sc1umax5_0}.
By relaxing the integer constraints (\ref{sc1umax5_0}e) with $x_{l}\in[0,1]$, problem \eqref{sc1umax5_0} becomes a convex problem.
For problem \eqref{sc1umax5_0} with relaxed constraints, the optimal solution can be obtained through the dual method \cite{boyd2004convex}.
We show that the dual method obtains the integer solution, which guarantees both optimality and feasibility of the original problem.
To obtain the optimal solution of problem (\ref{sc1umax5_0}), we have the following theorem.
\begin{theorem}
For problem (\ref{sc1umax5_0}), the optimal RIS on-off vector $\boldsymbol x$ and auxiliary variables $(y,\boldsymbol z)$ can be respectively expressed as:
\begin{equation}\label{sc1umax5_0eq1}
x_l=\left\{ \begin{array}{ll}
\!\!1, &\text{if}\; C_l>0, \\
\!\!0, &\text{otherwise},
\end{array} \right.
\end{equation}
\begin{equation}\label{sc1umax5_0eq2}
y=\left.\left(\frac{B}{(\ln2)\alpha}-\frac{\sigma^2}{p_1}\right)\right|_{\left(2^{\frac{R_1}{B}}-1\right)\frac{\sigma^2}{p_1}},
\end{equation}
and
\begin{equation}\label{sc1umax5_0eq3}
z_{lm}=\left\{ \begin{array}{ll}
\!\!1, &\text{if}\;  \alpha D_{lm}+\kappa_{1lm}-\kappa_{2lm}-\kappa_{3lm}>0, \\
\!\!0, &\text{otherwise},
\end{array} \right.
\end{equation}
where
\begin{equation}\label{sc1umax5_0eq3_1}
 C_{l}\!=\!\left\{ \begin{array}{l}
\!\!\!-\lambda N_1P_{\text{R}}+\alpha D_1+\sum_{m=2}^{L} (\kappa_{3ml}-\kappa_{1ml}), \quad  \text{if}\; l =1,\\
\!\!\! -\lambda N_lP_{\text{R}}+\alpha D_l+ \sum_{m=1}^{l-1} (\kappa_{2lm}-\kappa_{1lm})+\sum_{m=l+1}^{L} (\kappa_{3ml}-\kappa_{1ml}),  \quad \text{if}\; 2\leq  l \leq L-1,\\
\!\!\!-\lambda N_LP_{\text{R}}+\alpha D_L+ \sum_{m=1}^{L-1} (\kappa_{2lm}-\kappa_{1lm}),  \quad\text{if}\; l=L,
\end{array} \right.
\end{equation}
$\{\alpha,\kappa_{1lm},\kappa_{2lm},\kappa_{3lm}\}$ are the Lagrange multipliers associated with corresponding constraints of problem (\ref{sc1umax5_0}),
and $a_b=\max\{a,b\}$.
\end{theorem}

\itshape {Proof:}  \upshape
The dual problem of problem (\ref{sc1umax5_0}) with relaxed constraints is:
\begin{equation}\label{dueq1}
\mathop{\min}_{  \alpha, \boldsymbol \kappa}\quad \mathcal D( \alpha, \boldsymbol \kappa),
\end{equation}
where
\begin{equation}\label{dueq1_2}
\mathcal D( \alpha, \boldsymbol \kappa)=\left\{ \begin{array}{ll}
\!\!\!\mathop{\max}\limits_{  \boldsymbol x,y,\boldsymbol z}
&  \mathcal L (\boldsymbol x,y,\boldsymbol z, \alpha, \boldsymbol \kappa)
\\
\!\!\rm{s.t.}& y\geq \left(2^{\frac{R_1}{B}}-1\right)\frac{\sigma^2}{p_1},\\
&0\leq z_{lm}\leq1, \quad \forall l=2,\cdots, L, m=1,\cdots,l-1,\\
& 0\leq x_l\leq 1, \quad \forall  l \in\mathcal L,
\end{array} \right.
\end{equation}
\begin{align}\label{dueq2}
\mathcal L (\boldsymbol x,y,&\boldsymbol z, \alpha, \boldsymbol \kappa)={B  \log_2\left( 1+\frac{p_1y}{\sigma^2}\right)}-\lambda\left({ \mu p_1+P_1+P_{\text{B}}+\sum_{l=1}^L x_lN_lP_{\text{R}}}\right)
\nonumber\\ & +\alpha\left(D_{0}+\sum_{l=1}^L D_{l}x_l+ \sum_{l=2}^L \sum_{m=1}^{l-1}D_{lm} z_{lm}-y\right)
\nonumber\\&
+\sum_{l=2}^{L} \sum_{m=1}^{l-1}  \left[\kappa_{1lm}(z_{lm}-x_l-x_m+1)
+ \kappa_{2lm}(x_l -z_{lm})
+   \kappa_{3lm}(x_m -z_{lm})\right],
\end{align}
and $\boldsymbol \kappa=\{\kappa_{1lm},\kappa_{2lm},\kappa_{3lm}\}_{l=2,\cdots, L, m=1,\cdots,l-1}$.

To maximize the objective function in (\ref{dueq1_2}), which is a linear combination of $x_l$ and $z_{lm}$, we must
let the positive coefficients corresponding to the $x_l$ and $z_{lm}$  be 1.
Therefore, the optimal $x_l$ and $z_{lm}$ are thus given as \eqref{sc1umax5_0eq1} and \eqref{sc1umax5_0eq3}, respectively.

To optimize $y$ from \eqref{dueq1_2}, we set the first derivative of objective function to zero, i.e.,
\begin{equation}\label{sc1umax5_0eq5}
\frac{\partial\mathcal L (\boldsymbol x,y,\boldsymbol z, \alpha, \boldsymbol \kappa)}
{\partial y}=\frac{Bp_1}{(p_1y+\sigma^2)\ln2}-\alpha=0,
\end{equation}
which yields $y=\frac{B}{(\ln2)\alpha}-\frac{\sigma^2}{p_1}$.
Considering constraint (\ref{sc1umax5_0}c), we obtain the optimal solution to problem (\ref{sc1umax5_0}) as \eqref{sc1umax5_0eq2}.
 \hfill $\Box$

Theorem 2 states that RIS $l$ that has a positive coefficient $C_l$ should be on.
According to the expression of coefficient $C_l$ in \eqref{sc1umax5_0eq3_1}, the negative term $-\lambda N_lP_{\text{R}}$, is the effect of introducing additional power consumption if  RIS $l$ is on, while the remaining term represents the benefit of increasing the transmit rate by keeping RIS $l$ in operation.
When $C_l>0$, the benefit of increasing the transmit rate is larger than the effect of introducing additional power consumption,
which means that the energy efficiency can be improved if RIS $l$ is on.

The values of $(\alpha,\boldsymbol \kappa)$ are determined by the sub-gradient method \cite{bertsekas2009convex,8543183,9013160,yang2019energy}, which can be given by
\begin{eqnarray}
&&\!\!\!\!\!\!\!\!\!\!\!\!\!\!\!\!\!\!
\alpha=\left[ \alpha - \phi   \left(D_{0}+\sum_{l=1}^L D_{l}x_l+ \sum_{l=2}^L \sum_{m=1}^{l-1}D_{lm} z_{lm}-y\right)\right]^+,\label{sc1umax5_0eq6}\\
&&\!\!\!\!\!\!\!\!\!\!\!\!\!\!\!\!\!\!
\kappa_{1lm}=[\kappa_{1lm}-\phi(z_{lm}-x_l-x_m+1)]^+,\\
&&\!\!\!\!\!\!\!\!\!\!\!\!\!\!\!\!\!\!
\kappa_{2lm}=\left[\kappa_{2lm}- \phi(x_l -z_{lm})\right]^+,\\
&&\!\!\!\!\!\!\!\!\!\!\!\!\!\!\!\!\!\!
\kappa_{3lm}=\left[\kappa_{3lm}- \phi(x_m -z_{lm})\right]^+,\label{sc1umax5_0eq7}
\end{eqnarray}
where $\phi>0$ is a dynamically chosen step-size sequence and $[a]^+=\max(a,0)$.

 \begin{algorithm}[h]
\caption{  Dual Method for Problem (\ref{sc1umax5_1}) }
\begin{algorithmic}[1]\label{singleoptonoffAlo}
\STATE Initialize parameter $\lambda$ and set the accuracy $\epsilon$.
\REPEAT
\STATE Initialize dual variables $(\alpha,\boldsymbol \kappa)$.
\REPEAT
\STATE Update the RIS on-off vector $\boldsymbol x$ and the auxiliary variables $(y,\boldsymbol z)$  according to (\ref{sc1umax5_0eq1})-(\ref{sc1umax5_0eq3}).
\STATE Update dual variables $(\alpha,\boldsymbol \kappa)$ based on (\ref{sc1umax5_0eq6})-(\ref{sc1umax5_0eq7}).
\UNTIL the objective value (\ref{sc1umax5_0}) converges
\STATE Denote the objective value (\ref{sc1umax5_0}) by $H(\lambda)$.
\STATE Update $\lambda=\frac{B  \log_2\left( 1+\frac{p_1y}{\sigma^2}\right)}{ \mu p_1+P_1+P_{\text{B}}+\sum_{l=1}^L x_lN_lP_{\text{R}}}$.
\UNTIL $H(\lambda)<\epsilon$.
\end{algorithmic}
\end{algorithm}

By iteratively optimizing primal variables $(\boldsymbol x,y,\boldsymbol z)$ and  dual variables $(\alpha,\boldsymbol\kappa)$, the optimal RIS on-off vector is obtained.
The dual method for solving problem (\ref{sc1umax5_0}) and the Dinkelbach method to update parameter $\lambda$  are given in Algorithm \ref{singleoptonoffAlo}.
Notice that the optimal $x_l$ is either 0 or 1 according to \eqref{sc1umax5_0eq1}, even though
we relax $x_{l}$ as (\ref{dueq1_2}). Consequently, the optimal solution to problem (\ref{sc1umax5_0}) is obtained by using the dual method, i.e., $H(\lambda)$ in \eqref{equvialentProblem} is obtained for given $\lambda$.
Using the Dinkelbach method, we can obtain the root of $H(\lambda)=0$, which indicates that the optimal solution of fractional energy efficiency optimization problem (\ref{sc1umax5_1}) is obtained.



%
%
%

\subsection{Complexity Analysis}
\begin{algorithm}[t]
\caption{Iterative Optimization for Problem \eqref{sc1umax2}}
\begin{algorithmic}[1]\label{singleuseropiAlo}
\STATE Initialize $(\boldsymbol \theta^{(0)}, p_1^{(0)}, \boldsymbol x^{(0)})$. Set iteration number $n=1$.
\REPEAT
\STATE Given $\boldsymbol x^{(n-1)}$, solve the phase optimization problem \eqref{sc1umax3_2} by using Algorithm \ref{singlesosca} and the solution is denoted by $\boldsymbol \theta^{(n)}$.
\STATE Given $\boldsymbol x^{(n-1)}$ and the optimized $\boldsymbol \theta^{(n)}$, solve the power control problem \eqref{sc1umax3_5} according to Theorem 1 and the optimal power is denoted by $p_1^{(n)}$.
\STATE Given $(\boldsymbol \theta^{(n)},p_1^{(n)})$, solve the RIS on-off optimization problem \eqref{sc1umax5_1} by using Algorithm \ref{singleoptonoffAlo} and the solution is denoted by $\boldsymbol x^{(n)}$.
\STATE Set $n=n+1$.
\UNTIL the objective value (\ref{sc1umax2}) converges.
\end{algorithmic}
\end{algorithm}

The iterative algorithm for solving problem \eqref{sc1umax2} is given in Algorithm \ref{singleuseropiAlo}.
From Algorithm \ref{singleuseropiAlo}, the main complexity of solving problem \eqref{sc1umax2} lies in solving the phase optimization problem \eqref{sc1umax3_2} and the RIS on-off optimization problem \eqref{sc1umax5_1}.

According to Algorithm \ref{singlesosca}, to solve the phase optimization problem \eqref{sc1umax3_2},
the complexity lies in computing $\boldsymbol v^{(n)}=\text e^{-j\angle({\boldsymbol U}_{1}({\boldsymbol g}_{1} + {\boldsymbol U}_{1}^H {\boldsymbol v}^{(n-1)}) )}
$ at each iteration, which involves the complexity of $\mathcal O(QM)$.
Hence, the total complexity of solving problem \eqref{sc1umax3_2} with Algorithm \ref{singlesosca} is $\mathcal O(T_1QM)$, where $T_1$ is the total number of the iterations of Algorithm 1.

According to Algorithm \ref{singleoptonoffAlo}, the main complexity of solving problem \eqref{sc1umax5_1} lies in solving RIS on-off vector $\boldsymbol x$, which involves the complexity of $\mathcal O(L^2)$ based on \eqref{sc1umax5_0eq1} and \eqref{sc1umax5_0eq3_1}.
Hence, the complexity of solving problem \eqref{sc1umax5_1} with Algorithm \ref{singleoptonoffAlo} is $\mathcal O(T_2T_3L^2)$, where $T_2$ is the number of inner iterations by updating primal variables and dual variables and $T_3$ is the number of inner iterations by updating the parameter $\lambda$.

As a result, the total complexity of solving problem \eqref{sc1umax2} is $\mathcal O(T_0T_1QM+T_0T_2T_3L^2)$, where $T_0$ is the total number of iterations for Algorithm \ref{singleuseropiAlo}.
The complexity of the proposed Algorithm~\ref{singleuseropiAlo} grows quadratically with the number of all RISs and this complexity is lower than that of the SDR-based algorithm in \cite{8811733}.

\section{Energy Efficiency Optimization with Multiple Users}
In this section, we consider a general case with multiple users.
To solve the energy efficiency optimization problem in \eqref{sys2max1},
an iterative algorithm with low complexity is proposed via alternatingly optimizing the phase vector, beamforming vector, and RIS on-off vector.
\subsection{Phase Optimization}

Given beamforming vector $\boldsymbol w$ and RIS on-off vector $\boldsymbol x$, the total power consumption of the system is fixed and the energy efficiency maximization is equivalent to the sum-rate maximization.
Thus, given $(\boldsymbol w, \boldsymbol x)$, problem \eqref{sys2max1} reduces to:
\begin{subequations}\label{eemumax1}
\begin{align}
\mathop{\max}_{ \boldsymbol\theta,\boldsymbol \eta} \quad&  \sum_{k=1}^K \log_2\left(1+\eta_k\right) \tag{\theequation}  \\
\textrm{s.t.} \quad\:
&\eta_k\leq \frac{ \left|\left(\boldsymbol g_{k}^H+
\sum_{l=1}^L x_l\boldsymbol h_{kl}^H\boldsymbol\Theta_l \boldsymbol G_l\right) \boldsymbol w_k\right|^2}
{\sum_{i=1,i\neq k}^K   \left|\left(\boldsymbol g_{k}^H+
\sum_{l=1}^L x_l\boldsymbol h_{kl}^H\boldsymbol\Theta_l \boldsymbol G_l\right)\boldsymbol w_i\right|^2+\sigma^2}, \quad
\forall k\in\mathcal K,\\
&\eta_k\geq 2^{\frac{R_k}{B}}-1,\quad\forall k\in\mathcal K,\\
&\theta_{ln} \in[0,2\pi],   \quad  \forall l \in\mathcal L, n\in\mathcal N_l,
\end{align}
\end{subequations}
where $\boldsymbol \eta=[\eta_1,\cdots,\eta_K]^T$.
In \eqref{eemumax1},
$\boldsymbol \eta$ is a slack vector, which ensures that constraint (\ref{eemumax1}a) always holds with equality for the optimal solution.
Constraint (\ref{eemumax1}b) is added to guarantee the minimum rate requirement of each user.

Before optimizing $\boldsymbol \theta$,  we denote $s_{ln}=\text e^{j\theta_{ln}},\forall l \in\mathcal L, n\in\mathcal N_l$, $\boldsymbol s_l=[s_{l1},\cdots,s_{lN_l}]^T$, and $\boldsymbol s=[s_{11},\cdots,s_{1N_1},\cdots, s_{LN_L}]^T$.
With the help of $\boldsymbol s_l$, we show
$\boldsymbol h_{kl}^H \boldsymbol\Theta_l \boldsymbol G_l \boldsymbol w_i= \boldsymbol t_{kli}^H \boldsymbol s_l$, where $\boldsymbol t_{kli}=(\text{diag}(\boldsymbol h_{kl}^H ) \boldsymbol G_l \boldsymbol w_i)^* \in \mathbb C^{N_l}$.
Hence, constraint (\ref{eemumax1}a) can be rewritten as:
\begin{equation}\label{eemumax1eq0}
\eta_k\leq \frac{ \left|  {\boldsymbol t}_{kk}^H  {  \boldsymbol s}+ \bar{g}_{kk}\right|^2}
{\sum_{i=1,i\neq k}^K   \left|  {\boldsymbol t}_{ki}^H  {\boldsymbol s}  + \bar{g}_{ki}\right|^2+\sigma^2}, \quad
\forall k\in\mathcal K,
\end{equation}
where  $ {\boldsymbol t}_{ki}=[{\boldsymbol t}_{kl1};\cdots;{\boldsymbol t}_{klL}]$, $\bar{g}_{kk}= {\boldsymbol g}_{k}^H  \boldsymbol w_k$, and
$\bar{g}_{ki}= {\boldsymbol g}_{k}^H  \boldsymbol w_i$.

Problem \eqref{eemumax1} can be reformulated as:
\begin{subequations}\label{eemumax0}
\begin{align}
\mathop{\max}_{ \boldsymbol s,\boldsymbol \eta} \quad&  \sum_{k=1}^K \log_2\left(1+\eta_k\right) \tag{\theequation}  \\
\textrm{s.t.} \quad\:
&|s_{ln}|=1,   \quad  \forall l \in\mathcal L, n\in\mathcal N_l,\\
&(\ref{eemumax1eq0}),(\ref{eemumax1}b).
\end{align}
\end{subequations}
To handle the nonconvexity constraint (\ref{eemumax0}a), we use the penalty method and problem (\ref{eemumax0}) can be rewritten as:
\begin{subequations}\label{eemumax01}
\begin{align}
\mathop{\max}_{ \boldsymbol s,\boldsymbol \eta} \quad & \sum_{k=1}^K \log_2\left(1+\eta_k\right) +C\sum_{l=1}^L \sum_{n=1}^{N_l}(|s_{ln}|^2-1)
 \tag{\theequation}  \\
\textrm{s.t.} \quad\:
&|s_{ln}|\leq 1,   \quad  \forall l \in\mathcal L, n\in\mathcal N_l,\\
&(\ref{eemumax1eq0}),(\ref{eemumax1}b),
\end{align}
\end{subequations}
where $C$ is a large positive constant. Note that the penalty part $C\sum_{l=1}^L \sum_{n=1}^{N_l}(|s_{ln}|^2-1)$ enforces that $|s_{ln}|^2-1=0$ for the optimal solution of \eqref{eemumax01}.
 To solve the nonconvex problem in \eqref{eemumax01}, we utilize the SCA method.
The objective function of \eqref{eemumax01} can be approximated by:
\begin{equation}\label{eemumax1eq01}
\sum_{k=1}^K \log_2\left(1+\eta_k\right) +2C\sum_{l=1}^L \sum_{n=1}^{N_l}s_{ln}^{(n-1)} (s_{ln}-s_{ln}^{(n-1)}),
\end{equation}
where the second part is the first-order Taylor series of $C\sum_{l=1}^L \sum_{n=1}^{N_l}(|s_{ln}|^2-1)$ and the superscript $(n-1)$ means the value of the variable at the $(n-1)$-th iteration.
To handle the nonconvexity of constraint \eqref{eemumax1eq0}, we introduce variable $\beta_k$ and constraint \eqref{eemumax1eq0} is equivalent to:
\begin{equation}\label{eemumax1eq1}
 { \left|  {\boldsymbol t}_{kk}^H  {  \boldsymbol s}+ \bar{g}_{kk}\right|^2} \geq \beta_k \eta_k=\frac 1 4((\beta_k+\eta_k)^2-(\beta_k- \eta_k)^2),
\end{equation}
and
\begin{equation}\label{eemumax1eq1_2}
{\sum_{i=1,i\neq k}^K   \left|  {\boldsymbol t}_{ki}^H  {\boldsymbol s}  + \bar{g}_{ki}\right|^2+\sigma^2}\leq \beta_k,
 \end{equation}
where \eqref{eemumax1eq1_2} is convex, while it remains to  handle the nonconvexity of \eqref{eemumax1eq1}. We adopt an approximation of the difference of two convex functions (DC) and constraint \eqref{eemumax1eq1} can be approximated by:
\begin{align}\label{eemumax1eq1_3}
&\quad2\mathcal R(( {\boldsymbol t}_{kk}^H   {  \boldsymbol s}^{(n-1)}+\bar{g}_{kk})^H {\boldsymbol t}_{kk}^H  {  \boldsymbol s})-
{ \left|  {\boldsymbol t}_{kk}^H  {  \boldsymbol s}^{(n-1)}+ \bar{g}_{kk}\right|^2}
\nonumber\\&\geq \frac 1 4((\beta_k+\eta_k)^2-(\beta_k^{(n-1)}- \eta_k^{(n-1)})(\beta_k- \eta_k)+(\beta_k^{(n-1)}- \eta_k^{(n-1)})^2),
\end{align}
where the left hand side is the first-order Taylor expansion of ${ \left|  {\boldsymbol t}_{kk}^H  {  \boldsymbol s}+ \bar{g}_{kk}\right|^2}$ with respect to $\boldsymbol s$ at $\boldsymbol s=\boldsymbol s^{(n-1)}$. 

With the above approximations, the nonconvex problem in (\ref{eemumax01}) can be formulated in the following approximated convex problem:
\begin{subequations}\label{eemumax1_2}
\begin{align}
\mathop{\max}_{ {\boldsymbol   s},\boldsymbol  \eta,  \boldsymbol  \beta}\quad&  \sum_{k=1}^K \log_2\left(1+\eta_k\right) +2C\sum_{l=1}^L \sum_{n=1}^{N_l}s_{ln}^{(n-1)} (s_{ln}-s_{ln}^{(n-1)}),
 \tag{\theequation}\\
\textrm{s.t.}\quad \:
&\beta_k\geq0,\quad \forall k \in \mathcal K\\
&(\ref{eemumax1eq1_2}), (\ref{eemumax1eq1_3}), (\ref{eemumax01}a),
\end{align}
\end{subequations}
where $\boldsymbol \beta=[\beta_1,\cdots,\beta_K]^T$.
%
%
%
Problem (\ref{eemumax1}) can be solved by using the SCA method, where the approximated convex problem \eqref{eemumax1_2} is solved at each iteration.
The detailed process of using the SCA method to solve problem (\ref{eemumax1}) is analogous to Algorithm 1.
\subsection{Beamforming Optimization}
Given phase vector $\boldsymbol \theta$ and RIS on-off vector $\boldsymbol x$, problem \eqref{sys2max1} becomes:
\begin{subequations}\label{eemubeamax1}
\begin{align}
\mathop{\max}_{   \boldsymbol w, \boldsymbol \zeta} \quad& \frac{B\sum_{k=1}^K \log_2\left(1+\zeta_k\right)}
{\mu \boldsymbol w^H \boldsymbol w + \sum_{k=1}^K P_k +P_{\text{B}}+\sum_{l=1}^L x_l N_lP_{\text{R}}} \tag{\theequation}  \\
\textrm{s.t.} \quad\:
&\zeta_k \leq  \frac{ \left|\tilde{\boldsymbol g}_k^H \boldsymbol w_k\right|^2}
{\sum_{i=1,i\neq k}^K   \left|\tilde{\boldsymbol g}_k^H \boldsymbol w_i\right|^2+\sigma^2}, \quad
\forall k\in\mathcal K,\\
& \boldsymbol w^H \boldsymbol w \leq P_{\max},\\
&\zeta_k\geq 2^{\frac{R_k}{B}}-1,\quad\forall k\in\mathcal K,
\end{align}
\end{subequations}
where  $\boldsymbol \zeta=[\zeta_1,\cdots,\zeta_K]^T$ and $\tilde{\boldsymbol g}_k=\boldsymbol g_{k}+
\sum_{l=1}^Lx_l\boldsymbol G_l^H  \boldsymbol\Theta_l^H
\boldsymbol h_{kl}$.
In \eqref{eemubeamax1}, $\boldsymbol \zeta$ is a slack variable, which ensures that constraint (\ref{eemubeamax1}a) always holds with equality for the optimal solution.

To handle the nonconvexity of constraint  (\ref{eemubeamax1}a), we introduce a slack variable $\gamma_k>0$
and reformulate constraints (\ref{eemubeamax1}a) into the following equivalent form:
\begin{equation}\label{miso1min2eq1}
|\tilde{\boldsymbol g}_k^H \boldsymbol w_k|^2\geq \gamma_k \zeta_k,
\end{equation}
and
\begin{equation}\label{miso1min2eq2}
\sum_{i=1,i\neq k}^K|\tilde{\boldsymbol g}_k^H \boldsymbol w_i|^2+\sigma^2 \leq \gamma_k.
\end{equation}

Without loss of generality, the term $\tilde{\boldsymbol g}_k^H \boldsymbol w_k$ in constraint \eqref{miso1min2eq1} can be expressed as a real number through an arbitrary rotation to beamforming $\boldsymbol w_k$.
As a result, constraint \eqref{miso1min2eq1} can be equivalent to
$
\mathcal R(\tilde{\boldsymbol g}_k^H \boldsymbol w_k)\geq \sqrt{\gamma_k \zeta_k}.
$
Replacing concave function $\sqrt{\gamma_k \zeta_k}$ with the first-order Taylor series, constraint \eqref{miso1min2eq1} becomes:
\begin{equation}\label{miso1min2eq1_3}
\mathcal R(\tilde{\boldsymbol g}_k^H \boldsymbol w_k)\geq \sqrt{\gamma_k^{(n-1)} \zeta_k^{(n-1)}}+\frac1 2 \sqrt{\frac{\gamma_k^{(n-1)}} {\zeta_k^{(n-1)}}} (\zeta_k-\zeta_k^{(n-1)})+\frac1 2 \sqrt{\frac{\zeta_k^{(n-1)}}{\gamma_k^{(n-1)}} } (\gamma_k-\gamma_k^{(n-1)}).
\end{equation}

Using the above approximations, the nonconvex problem in (\ref{eemubeamax1}) can be formulated in the following approximated problem:
\begin{subequations}\label{eemubeamax1_2}
\begin{align}
\mathop{\max}_{\boldsymbol w,\boldsymbol  \zeta,  \boldsymbol  \gamma}\quad&
 \frac{B\sum_{k=1}^K \log_2\left(1+\zeta\right)}
{\mu \boldsymbol w^H \boldsymbol w + \sum_{k=1}^K P_k +P_{\text{B}}+\sum_{l=1}^L x_l N_lP_{\text{R}}} \tag{\theequation}\\
\textrm{s.t.}\quad \:
&(\ref{miso1min2eq2}),(\ref{miso1min2eq1_3}), (\ref{eemubeamax1}b), (\ref{eemubeamax1}c),\\
&\gamma_k\geq0,\quad \forall k \in \mathcal K.
\end{align}
\end{subequations}
Since the objective function is a concave function divided by a convex function and the feasible set is convex, the optimal solution of problem \eqref{eemubeamax1_2} can be obtained by the Dinkelbach method in \cite{dinkelbach1967nonlinear}.
As a result, the original beamfoming optimization problem in (\ref{eemubeamax1}) can be solved by using the SCA method, where \eqref{eemubeamax1_2} is solved optimally by using the Dinkelbach method at each iteration.
%
%
%

\subsection{RIS On-Off Optimization}
\begin{algorithm}[t]
\caption{Greedy Method For RIS On-Off Optimization}
\small
\begin{algorithmic}[1]\label{multipleusergreedy}
\STATE Initialize $\mathcal A=\{1,\cdots,L\}$ and $x_l=1$, $\forall l\in\mathcal A$.
\STATE Calculate the objective value \eqref{sys2max1}, which is denoted by $E_0$.
\WHILE {$\mathcal A \neq \emptyset$}
\FOR{$l\in\mathcal A$}
\STATE Turn off RIS $l$, i.e., construct a new RIS on-off solution, $x_l=0$, $x_m=1$, $x_n=0$, $\forall m \in\mathcal A\setminus\{l\}$,  $n \in\mathcal L\setminus \mathcal A$.
\STATE If the new RIS on-off solution is feasible, calculate the  objective value \eqref{sys2max1}, which is denoted  by $E_l$.
\STATE If the new RIS on-off solution is infeasible, set $E_l=0$.
\ENDFOR
\STATE Calculate $k=\arg\max_{j\in\mathcal A\cup\{0\}} E_j$.
\IF{$k\neq 0$}
\STATE Set $\mathcal A=\mathcal A \setminus\{k\}$ and $E_0=E_k$.
\ELSE
\STATE Break and jump to Step 16.
\ENDIF
\ENDWHILE
\STATE Output $x_l=1$, $x_m=0$,  $\forall l\in\mathcal A$, $m\in\mathcal L\setminus\mathcal A$.
\end{algorithmic}
\end{algorithm}

Given phase vector $\boldsymbol \theta$ and beamforming vector $\boldsymbol w$, problem \eqref{sys2max1} is a nonlinear integer optimization problem with respect to the RIS on-off vector $\boldsymbol x$.
Since the nonlinear integer optimization problem is NP-hard in general, it is hard to obtain the globally optimal solution with polynomial complexity.
To tackle this difficulty, we use the greedy method to solve the RIS on-off optimization problem.

To solve problem \eqref{sys2max1} with fixed $(\boldsymbol \theta,\boldsymbol w)$, we propose a greedy method based algorithm to optimize the integer RIS on-off vector, which is given in
Algorithm \ref{multipleusergreedy}.
The basic idea of Algorithm \ref{multipleusergreedy} is that we try to turn off one RIS at each time if the objective value \eqref{sys2max1} can be improved and the new solution is feasible.

At step 1 of Algorithm \ref{multipleusergreedy}, set $\mathcal A$ is the set of active RISs that are serving users.
At step~2, $E_0$ is the initial objective value.
Step 5 means that we construct a new solution by turning off one RIS from $\mathcal A$.
If the new solution is feasible, we calculate the objective value \eqref{sys2max1}, as shown at step 6.
Otherwise, the new solution is infeasible, we denote the objective value as zero at step 7.
At step 9, we compare the energy efficiency values of these new solutions and the initial solution.
If $k\neq0$, the index value $k$ means that turning off one RIS can increase the energy efficiency and turning off RIS $k$ leads to the highest energy efficiency compared to turning off any other RIS $l\neq k$.
In this case, it is energy efficient to turn off RIS $k$, then we subtract index $k$ from the active set $\mathcal A$ and update the initial energy efficiency value for the next iteration, as shown at step 11.
If $k=0$, it means that turning off any one RIS can lead to low energy efficiency.
In this case, turning off any one RIS cannot increase the energy efficiency, which means that we need to terminate the loop and output the active set.

Since the objective value of problem (\ref{sys2max1}) is increasing at each iteration and the objective value always has a finite upper bound,
 Algorithm~\ref{multipleusergreedy} must converge.

\subsection{Complexity Analysis}
In summary, the iterative algorithm for solving the general multi-user energy efficiency maximization problem in \eqref{sys2max1} is given in Algorithm \ref{multipleuseropiAlo}.
From Algorithm \ref{multipleuseropiAlo}, the complexity of solving problem \eqref{sys2max1} is dominated by the complexity of solving the phase optimization problem in \eqref{eemumax1}, beamforming optimization problem \eqref{eemubeamax1}, and optimizing the RIS on-off status.

The phase optimization problem \eqref{eemumax1} is solved by using the SCA method.
Since there are $3K$ constraints in problem (\ref{eemumax1}), the number of iterations that are required for SCA method is $\mathcal O(\sqrt{2K+Q}\log_2(1/\epsilon_1))$ \cite{grant2008cvx}, where $\epsilon_1$ is the accuracy of the SCA method for solving problem (\ref{eemumax1}).
At each iteration, the complexity of solving  problem \eqref{eemumax1_2} is $\mathcal O (S_1^2 S_2)$, where
$S_1=2K+Q$ is the total number of variables and
$S_2=3K+Q$ is the total number of constraints \cite{lobo1998applications}.
Thus, the total complexity of the SCA method for solving problem (\ref{eemumax1}) is
$\mathcal O(K^{3.5}\log_2(1/\epsilon_1))$.
With similar analysis, the  total complexity of the SCA method for solving the beamforming optimization problem \eqref{eemubeamax1} is
$\mathcal O(TK^{3.5}\log_2(1/\epsilon_2))$, where $T$ is the number of iterations for solving problem \eqref{eemubeamax1_2} with the Dinkelbach method and $\epsilon_2$ is the accuracy of the SCA method for solving problem (\ref{eemubeamax1}).
According to Algorithm \ref{multipleusergreedy}, the main complexity of the RIS on-off optimization lies in calculating the objective value \eqref{sys2max1}, which involves the complexity of $\mathcal O(QM)$.
Thus, the total complexity of  Algorithm \ref{multipleusergreedy} is $\mathcal O(L^2QM)$, where $\mathcal O(L^2)$ is the total number of iterations for Algorithm \ref{multipleusergreedy}.

As a result, the total complexity of Algorithm \ref{multipleuseropiAlo} for solving problem \eqref{sys2max1} is $\mathcal O(S_3K^{3.5}\log_2(1/\epsilon_1)+S_3TK^{3.5}\log_2(1/\epsilon_2)+S_3 L^2QM)$, where $S_3$ is the number of iterations for Algorithm \ref{multipleuseropiAlo}.
The proposed Algorithm \ref{multipleuseropiAlo} has a lower complexity than the SDR-based algorithm in \cite{8811733}.

\begin{algorithm}[t]
\caption{Iterative Optimization for Problem \eqref{sys2max1}}
\begin{algorithmic}[1]\label{multipleuseropiAlo}
\STATE Initialize $(\boldsymbol \theta^{(0)}, \boldsymbol w^{(0)}, \boldsymbol x^{(0)})$. Set iteration number $n=1$.
\REPEAT
\STATE Given $(\boldsymbol w^{(n-1)}, \boldsymbol x^{(n-1)})$, solve the phase optimization problem \eqref{eemumax1} by using the SCA method in Section IV-A and the solution is denoted by $\boldsymbol \theta^{(n)}$.
\STATE Given $(\boldsymbol \theta^{(n)}, \boldsymbol x^{(n-1)})$,  solve the beamforming optimization problem \eqref{eemubeamax1} by using the SCA method in Section IV-B and the solution is denoted by $\boldsymbol w^{(n)}$.
\STATE Given $(\boldsymbol \theta^{(n)}, \boldsymbol w^{(n)})$, optimize the RIS on-off vector by using Algorithm \ref{multipleusergreedy} and the solution is denoted by $\boldsymbol x^{(n)}$.
\STATE Set $n=n+1$.
\UNTIL the objective value (\ref{sys2max1}) converges.
\end{algorithmic}
\end{algorithm}

\section{Simulation Results and Analysis}
There are $K$ users  uniformly distributed in a square area of size $300$ m $\times$ $300$~m  with the BS located at its center.
There are $L$ RISs and the location of RIS $l$ is given by $(\cos(2l\pi/L), \sin(2l\pi/L))\times100$ m.
The main system parameters are listed in
Table I \cite{4786499}.
The relay is assumed to transmit with the maximum power $P_{\text T}$.
Unless specified otherwise, we choose a maximum transmit power BS $P_{\max}=50$ dBm for the BS, a total of $M=8$ BS transmit antennas,  a total of $L=8$ RISs, a penalty factor $C=10^3$, $K=1$,  an equal
number of reflecting elements $N_1=\cdots=N_L=N=4$, and an equal rate demand $R_1=\cdots=R_K=R=1$ Mbps (i.e., 1 bps/Hz).
We compare the proposed scheme using distributed RISs (labeled `DRIS') with the following schemes: the conventional scheme with the central deployment of one RIS located at (100, 0) m in \cite{8741198} (labeled `CRIS') and the conventional AF relay scheme \cite{6671453} (labeled `AFR').
In particular, the number of reflecting elements for one central RIS in CRIS is set as the total number of reflecting elements for all RISs in DRIS.
In AFR, we consider the same deployment of DRIS, i.e., there are $L$ AF relays, where AF $l$ with $N$ antennas is located at $(\cos(2l\pi/L), \sin(2l\pi/L))\times100$ m.
\begin{table}[t]
\centering
\caption{System  Parameters} \label{tab:complexity}
\begin{tabular}{ccc}
  \hline
  \hline
  Parameters &   Values \\ \hline
Bandwidth of the BS  $B$ &1 MHz  \\
Noise power  $\sigma^2$& $-104$ dBm \\
Maximum transmit power of the AF relay  $P_{\text T}$ & 30 dBm\\
Small scale fading model, $\forall k,l,m,n$& $[\boldsymbol g_{k}]_{m}, [\boldsymbol h_{kl}]_{m}, [\boldsymbol G_l]_{mn} \sim\mathcal {CN}(0,1)$\\
Large scale fading model at distance $d$ & $\frac{10^{-3.53}}{d^{3.76}}$\\
Circuit power of the BS $P_{\text{B}}$ & 39 dBm\\
Power amplifier efficiency  at the BS/ AF relay $\nu$  & 0.8\\
Circuit power of each user $P_k$ & 10 dBm\\
Circuit power of each RIS element $P_{\text{R}}$ & 10 dBm\\
Circuit power of each AF relay transmit-receive antenna $P_{\text{A}}$ & 10 dBm\\
  \hline
  \hline
\end{tabular}
\end{table}

\begin{figure}[t]
\centering
\includegraphics[width=3.9in]{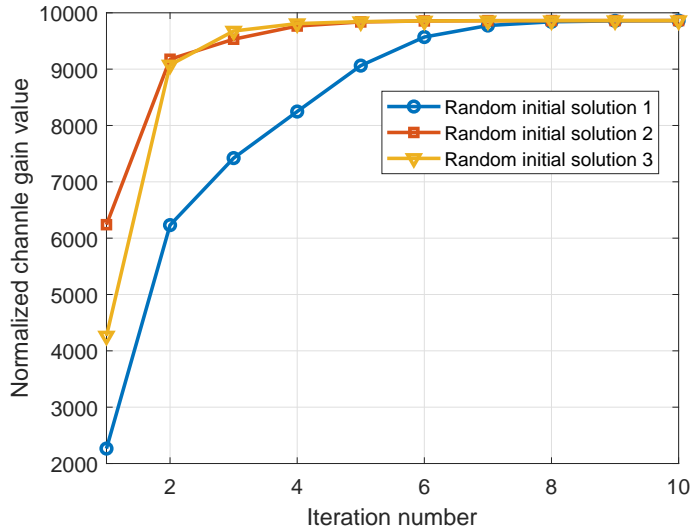}
\vspace{-1.5em}
\caption{Convergence behaviour of Algorithm \ref{singlesosca} with different initial solutions.} \label{fig1}
\vspace{-1.5em}
\end{figure}

Fig. \ref{fig1} illustrates the convergence of Algorithm \ref{singlesosca} using different initial solutions.
In this figure, the normalized channel gain value means the channel gain divided by the noise power, i.e., $\frac{ \left| {\boldsymbol g}_{1} + {\boldsymbol U}_{1}^H {\boldsymbol v} \right|^2 } {\sigma^2}$.
It can be seen that the proposed algorithm converges fast, and eight iterations
are sufficient to converge, which shows the effectiveness of the proposed algorithm.

\begin{figure}[t]
\centering
\includegraphics[width=3.9in]{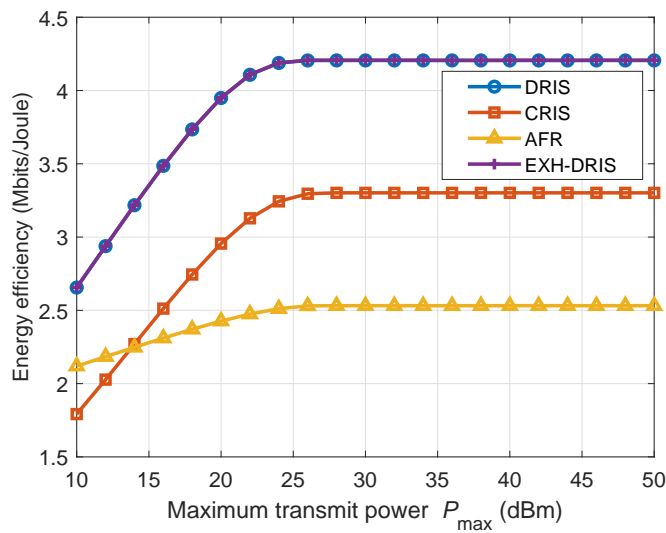}
\vspace{-1.5em}
\caption{Energy efficiency versus the maximum transmit power $P_{\max}$ of the BS.} \label{fig2}
\vspace{-1.5em}
\end{figure}

Fig. \ref{fig2} shows how the energy efficiency changes as the maximum transmit power of the BS varies.
In this figure, the EXH-DRIS scheme is an exhaustive search method that can find a near optimal solution of
problem (\ref{sc1umax2}). Hereinafter, the  EXH-DRIS scheme refers to the proposed DRIS algorithm with 1000 initial starting points.
In this simulation, EXH-DRIS can obtain  1000 solutions, and the solution with the highest energy efficiency is treated as the near optimal solution.
It is shown that the energy efficiency  of all schemes first increases and then remains stable as the maximum transmit power of the BS increases.
This is because energy efficiency is not a monotonically increasing function of the maximum transmit power, as shown in \eqref{sc1umax3_5eq1}.
When $P_{\max}\geq25$ dBm, the exceed transmit power is not used since it will decrease the energy efficiency.
 Fig. \ref{fig2} also shows that the proposed
DRIS scheme outperforms the CRIS and AFR schemes.
For high maximum transmit power of the BS,  DRIS can increase up to  27\% and 68\% energy efficiency compared to CRIS and AFR, respectively.
Moreover, the proposed DRIS scheme achieves almost the same performance as the EXH-DRIS scheme, which indicates
that the proposed DRIS can achieve the near optimum solution.

\begin{figure}[t]
\centering
\includegraphics[width=3.9in]{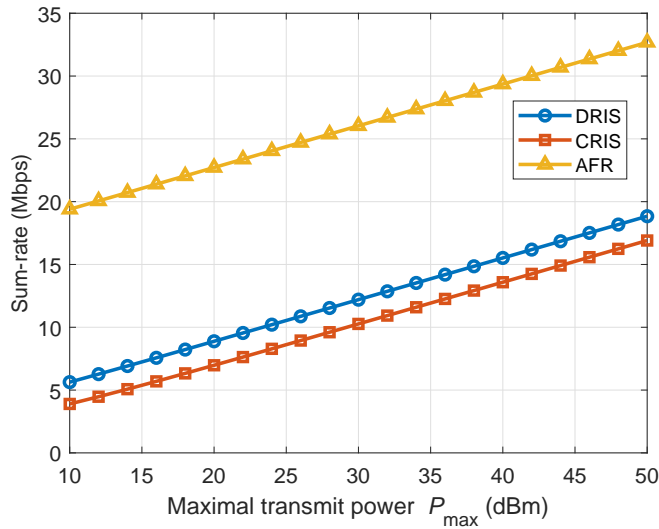}
\vspace{-1.5em}
\caption{Sum-rate versus the maximum transmit power $P_{\max}$ of the BS with $R=0$.} \label{fig3}
\vspace{-1.5em}
\end{figure}

Fig. \ref{fig3} shows how the sum-rate changes as the maximum transmit power of the BS varies.
It is found that the sum-rate of all schemes linearly increases with the logarithmic maximum transmit power of the BS.
We can see that AFR achieves the best performance.
This is because the AF relay is an active terminal by transmitting the received signal to the user, while the RIS is only a passive reflecting structure.
From Fig.~\ref{fig3},
DRIS can increase up to  26\% sum-rate compared to CRIS.
This is  due to the benefits of distributed deployment.
Multiple RISs are spatially distributed in DRIS, which can provide more than one path of received signal compared to CRIS with only one central RIS.

\begin{figure}[t]
\centering
\includegraphics[width=3.9in]{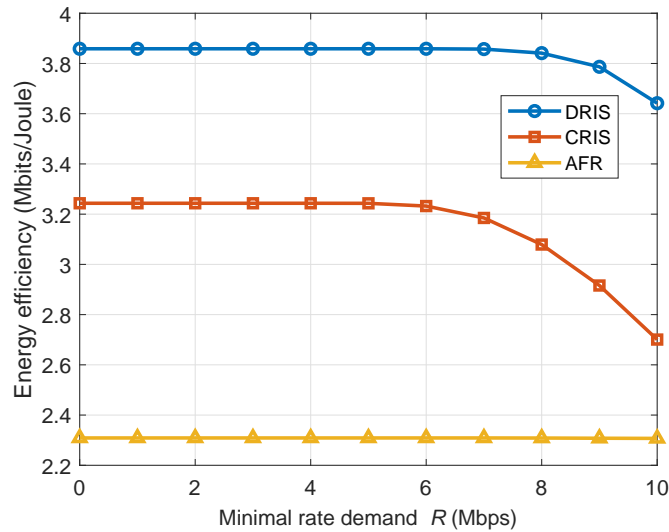}
\vspace{-1.5em}
\caption{Energy efficiency versus the minimum rate demand $R$.} \label{fig32}
\vspace{-1.5em}
\end{figure}

Fig. \ref{fig32} shows the energy efficiency versus the minimum rate demand.
From this figure, DRIS achieves the best performance.
In particular, DRIS can achieve up to 33\% and 67\% gains in terms of energy efficiency compared to CRIS and AFR, respectively.
Both DRIS and CRIS always achieve a better performance than AFR.
This is due to the fact that the power consumption for the RIS is much lower compared to the transmit power of the AF relay.
It is also found that DRIS achieves better performance than CRIS, which indicates the benefit of distributed deployment of RISs.
This is because DRIS exhibits a better spectrum efficiency compared to CRIS, and CRIS is more sensitive to high minimum rate demand than DRIS.
From Fig.~\ref{fig32}, we can observe that the energy efficiency remains stable  when minimum rate demand is low.
However, for a high minimum rate demand, the energy efficiency decreases rapidly for both CRIS and DRIS.
This is because a high minimum rate demand requires the BS to transmit with high power, which consequently degrades the energy efficiency performance.
Fig. \ref{fig32} also demonstrates that, as the minimum rate demand increases, the energy efficiency of CRIS decreases faster than DRIS.

\begin{figure}[t]
\centering
\includegraphics[width=3.9in]{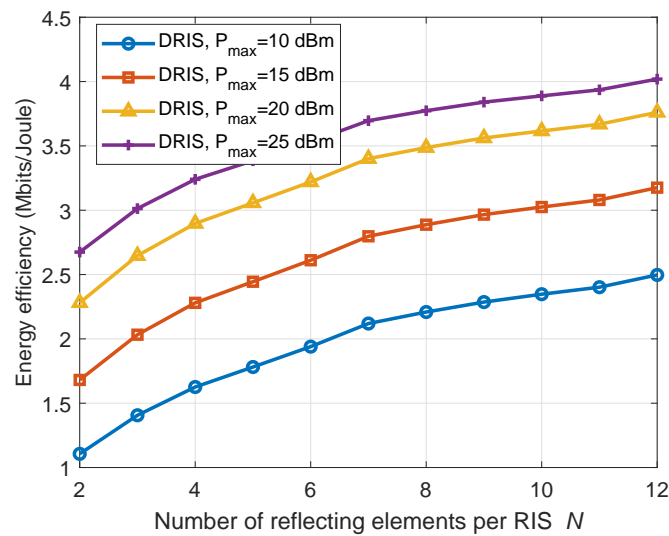}
\vspace{-1.5em}
\caption{Energy efficiency versus the number of reflecting elements $N$ for each RIS with $L=4$.} \label{fig5}
\vspace{-1.5em}
\end{figure}

\begin{figure}[t]
\centering
\includegraphics[width=3.9in]{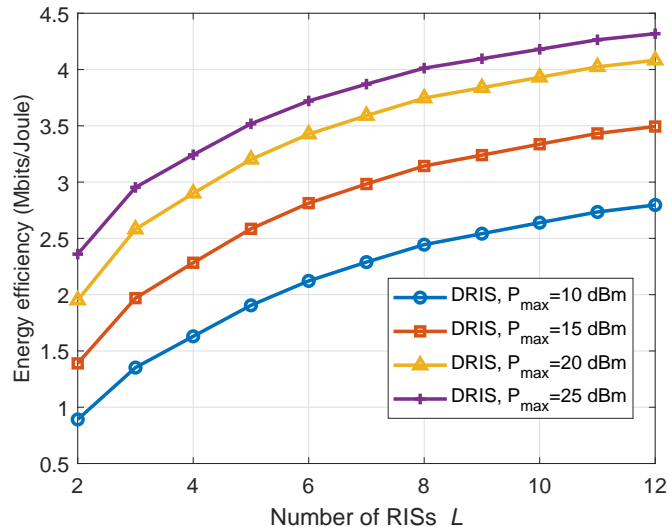}
\vspace{-1.5em}
\caption{Energy efficiency versus the number of RISs $L$ with $N=4$.} \label{fig6}
\vspace{-1.5em}
\end{figure}

Figs.~\ref{fig5} and \ref{fig6} show the energy efficiency versus the number of reflecting elements for each RIS and the number of RISs, respectively.
From these figures, we can see that the energy efficiency of DRIS monotonically increases with the number of reflecting elements and the number of RISs.
This is because large number of reflecting elements and RISs can lead to high spectral efficiency and the power consumption of introducing additional reflecting elements and RISs is low, which result in high energy efficiency of the system.
According to Figs.~\ref{fig5} and \ref{fig6}, it is also found that the energy efficiency of DRIS increases faster with the number of RISs than that with the number of reflecting elements for each RIS, which shows that it is more energy efficient to deploy with multiple RISs.
For the same total number of all reflecting elements, i.e., the same $NL$,
we consider the following two configurations: (a)  $N=12$, $L=4$, $P_{\max}=25$ dBm, and (b)  $N=4$, $L=12$, $P_{\max}=25$ dBm.
From Figs.~\ref{fig5} and \ref{fig6}, the energy efficiency of configuration (a) is 4.0 Mbits/Joule and the energy efficiency of configuration (b) is 4.3 Mbits/Joule,
which shows that the energy efficiency of $L>N$ is better than that of $L<N$.

\begin{figure}[t]
\centering
\includegraphics[width=3.9in]{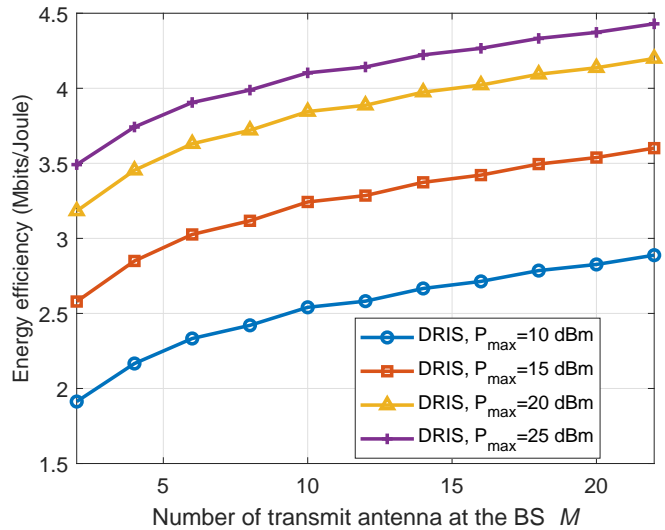}
\vspace{-1.5em}
\caption{Energy efficiency versus the number of transmit antennas $M$ at the BS.} \label{fig7}
\vspace{-1.5em}
\end{figure}

In Fig. \ref{fig7}, we show the energy efficiency versus the number of transmit antennas at the BS with various maximum transmit power of the BS.
Fig. \ref{fig7} demonstrates that the energy efficiency increases rapidly for a small number of transmit antennas at the BS, however, this increase becomes slower for a larger number of transmit antennas at the BS.
This is because a high number of transmit antennas at the BS leads to high power consumption, which consequently decreases the slope of increase of the energy efficiency.
From Fig. \ref{fig7}, we can also see that the energy efficiency increases with the maximum transmit power of the BS for a given number of transmit antenna at the BS.
Moreover, for high value of the BS maximum transmit power, the energy efficiency slowly increases with the maximum transmit power.

\begin{figure}[t]
\centering
\includegraphics[width=3.9in]{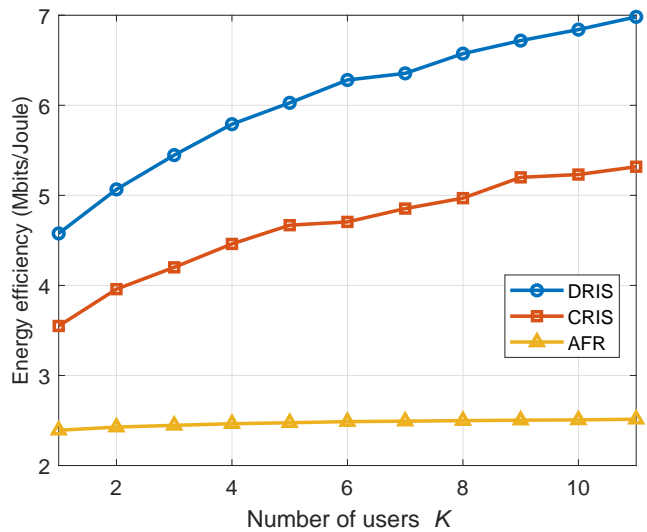}
\vspace{-1.5em}
\caption{Energy efficiency versus the number of users.} \label{fig8}
\vspace{-1.5em}
\end{figure}

The energy efficiency versus the number of users is shown in Fig.~\ref{fig8}.
From Fig.~\ref{fig8}, we observe that the energy efficiency of AFR is always low due to the fact that the transmit power of the AF relay is high.
Clearly, the proposed DRIS is always better than CRIS and AFR especially when the number of users is large.

\section{Conclusion}

In this paper, we have investigated the resource allocation problem for a wireless communication network with distributed RISs.
The RIS phase shifts, BS transmit beamforming, and RIS on-off status were jointly optimized to maximize the system energy efficiency while
satisfying minimum rate demand, maximum transmit power, and unit-modulus constraints.
To solve this problem, we have proposed two iterative algorithms with low complexity for the  single-user case and multi-user case, respectively.
In particular, the phase optimization problem was solved by using the SCA method, where the closed-form solution was obtained at each step for the single-user case.
Numerical results have shown that the proposed scheme outperforms conventional schemes in terms of energy efficiency, especially for small maximum transmit power and large number of users.

\bibliographystyle{IEEEtran}
\bibliography{IEEEabrv,ref}

\end{document}